\RequirePackage[hyphens]{url}
\RequirePackage[prologue,table]{xcolor}

\documentclass[manuscript,screen,authorversion,nonacm]{acmart}

\pdfoutput=1                   
\usepackage{graphicx}
\graphicspath{{./}, {figures/}}
\DeclareGraphicsExtensions{.pdf,.jpeg,.png}

\usepackage{amsmath}
\usepackage{amssymb}           \usepackage{amsthm}
\usepackage{balance}           \usepackage{booktabs}          \usepackage{caption}           \usepackage{enumitem}
\usepackage{textcomp}          \usepackage{xspace}            \usepackage[hyphens]{url}

\usepackage{lipsum}
\usepackage{subcaption}        
\usepackage{color}
\usepackage[prologue,table]{xcolor}

\usepackage{mscommon}
\newcommand{\inline}[1]{{\small\fontfamily{cmtt}\selectfont{#1}}}

\usepackage{varioref}          \usepackage{hyperref}          \usepackage[capitalise]{cleveref}  
\newcommand{\mytitle}{Meeting in the notebook: a notebook-based environment for micro-submissions in data science collaborations}
\newcommand{\mytitleshort}{A notebook-based environment for micro-submissions in data science collaborations}

\newcommand{\Assemble}{Assembl\'{e}\xspace}
\newcommand{\Ballet}{Ballet\xspace}
\newcommand{\assembleurl}{https://github.com/ballet/ballet-assemble}
\newcommand{\gatewayurl}{https://github.com/ballet/github-oauth-gateway}
\newcommand{\censusurl}{https://github.com/HDI-Project/ballet-predict-census-income}

\makeatletter
\if@ACM@anonymous
  \renewcommand{\assembleurl}{https://anonymous.4open.science/r/fcd5a2ee-ad0d-493e-bedb-6c4c465bb3b9/}
  \renewcommand{\gatewayurl}{https://anonymous.4open.science/r/f0d659ed-0c51-45de-935f-9657ed0fab4f/}
  \newcommand{\censusurl}{https://anonymous.4open.science/r/032af28b-f254-44d0-9018-cf32af43948b/}
\fi
\makeatother

\setcopyright{acmcopyright}
\copyrightyear{2018}
\acmYear{2018}
\acmDOI{10.1145/1122445.1122456}
\acmConference[Woodstock '18]{Woodstock '18: ACM Symposium on Neural
  Gaze Detection}{June 03--05, 2018}{Woodstock, NY}
\acmBooktitle{Woodstock '18: ACM Symposium on Neural Gaze Detection,
  June 03--05, 2018, Woodstock, NY}
\acmPrice{15.00}
\acmISBN{978-1-4503-XXXX-X/18/06}

\begin{document}

\title[\mytitleshort]{\mytitle}

\author{Micah J. Smith}
\email{micahs@mit.edu}
\affiliation{  \institution{Massachusetts Institute of Technology}
  \city{Cambridge}
  \state{Massachusetts}
  \country{USA}
}

\author{J\"{u}rgen Cito}
\email{juergen.cito@tuwien.ac.at}
\affiliation{  \institution{TU Wien}
  \city{Vienna}
  \country{Austria}
}
\affiliation{  \institution{Massachusetts Institute of Technology}
  \city{Cambridge}
  \state{Massachusetts}
  \country{USA}
}

\author{Kalyan Veeramachaneni}
\email{kalyanv@mit.edu}
\affiliation{  \institution{Massachusetts Institute of Technology}
  \city{Cambridge}
  \state{Massachusetts}
  \country{USA}
}

\begin{abstract}
    Developers in data science and other domains frequently use computational notebooks to create exploratory analyses and prototype models. However, they often struggle to incorporate existing software engineering tooling into these notebook-based workflows, leading to fragile development processes. We introduce \Assemble, a new development environment for collaborative data science projects, in which promising code fragments of data science pipelines can be contributed as pull requests to an upstream repository entirely from within JupyterLab, abstracting away low-level version control tool usage. We describe the design and implementation of \Assemble and report on a user study of 23 data scientists.
\end{abstract}


\maketitle

\section{Introduction}
\label{sec:intro}

Software engineering is a mature discipline with robust processes for team-based development, including version control tools, build tools, documentation generators, unit test frameworks, linters and formatters, and code review interfaces. Taken together, these tools address or prevent a wide class of issues in collaborative software development and make the resulting processes much more robust.
Although diverse types of practitioners write code, there has traditionally been a divide between the processes used by software engineers and those used by other developers, such as data scientists. Data scientists in particular often do exploratory data analysis and prototyping within a computational notebook, such as Jupyter Notebook or its successor, JupyterLab \cite{kery2017exploring}. These ``notebook-based'' developers then productionize completed analyses, either by themselves or together with specialized teammates. The exploratory code may be refactored, made more efficient and modular, or rewritten in a different framework or to target a different runtime. This two-stage development process presents challenges for developers that are not as fluent in using development tools, and makes the resulting software vulnerable to quality issues.

In response, researchers have proposed a variety of improvements to the data science development process. One recent idea is \Ballet, a software framework that supports collaborative data science development by composing a data science pipeline from a collection of modular patches that can be written in parallel \cite{smith2020enabling}. Within the larger practice of data science, \Ballet focuses on predictive machine learning (ML), and more specifically, feature engineering, a key subprocess in which raw variables are transformed into useful features suitable for learning. Thus developers might split up work to write individual feature definitions in short code snippets, while the \Ballet framework is responsible for structuring contributions, validating them in the context of the prediction task, and composing them into an executable feature engineering pipeline.

Typically, a developer contributing to a \Ballet project (or other kinds of data science projects) does exploratory work in a notebook before finally identifying a worthwhile patch to contribute. By this time, their notebook may be ``messy'' \cite{head2019managing} and the process to extract the relevant patch and translate it into a well-structured contribution to a shared repository becomes a challenge. Developers usually need to rely on a completely separate set of tools for this process, jettisoning the notebook for command line or GUI tools targeting team-based version control. This \emph{patch contribution task} is challenging even for developers experienced with open-source practices \cite{gousios2015work}, and is only more acute for data science developers who are less familiar with open-source development workflows.

To address this challenge, we propose a novel development environment, \Assemble.\footnote{\url{\assembleurl}}\textsuperscript{,}\footnote{\emph{\Assemble} is a ballet move that involves lifting off the floor on one leg and landing on two.}
\Assemble solves the patch contribution task to data science collaborations that use \Ballet\footnote{\Assemble targets contributions to \Ballet projects because of the structure that these projects impose on code contributions, but can be extended to support other settings as well.} by providing a higher-level interface for contributing code snippets within a larger notebook to an upstream repository --- meeting data science developers where they are most comfortable.
Rather than asking developers to productionize their exploratory notebooks, \Assemble enables data scientists to both develop and contribute back their code without leaving the notebook. A code fragment selected by a developer can be automatically formulated as a pull request (PR) to an upstream GitHub repository using an interface situated within the notebook environment itself, automating and abstracting away usage of low-level tools for testing and team-based development. It integrates tightly with Binder,\footnote{\url{https://mybinder.org/}} a community service for cloud-hosted notebooks, so that developers can get started with no setup required.

\begin{figure}
    \centering
    \includegraphics[page=1, clip=true, trim={10 20 60 20}, width=0.45\textwidth]{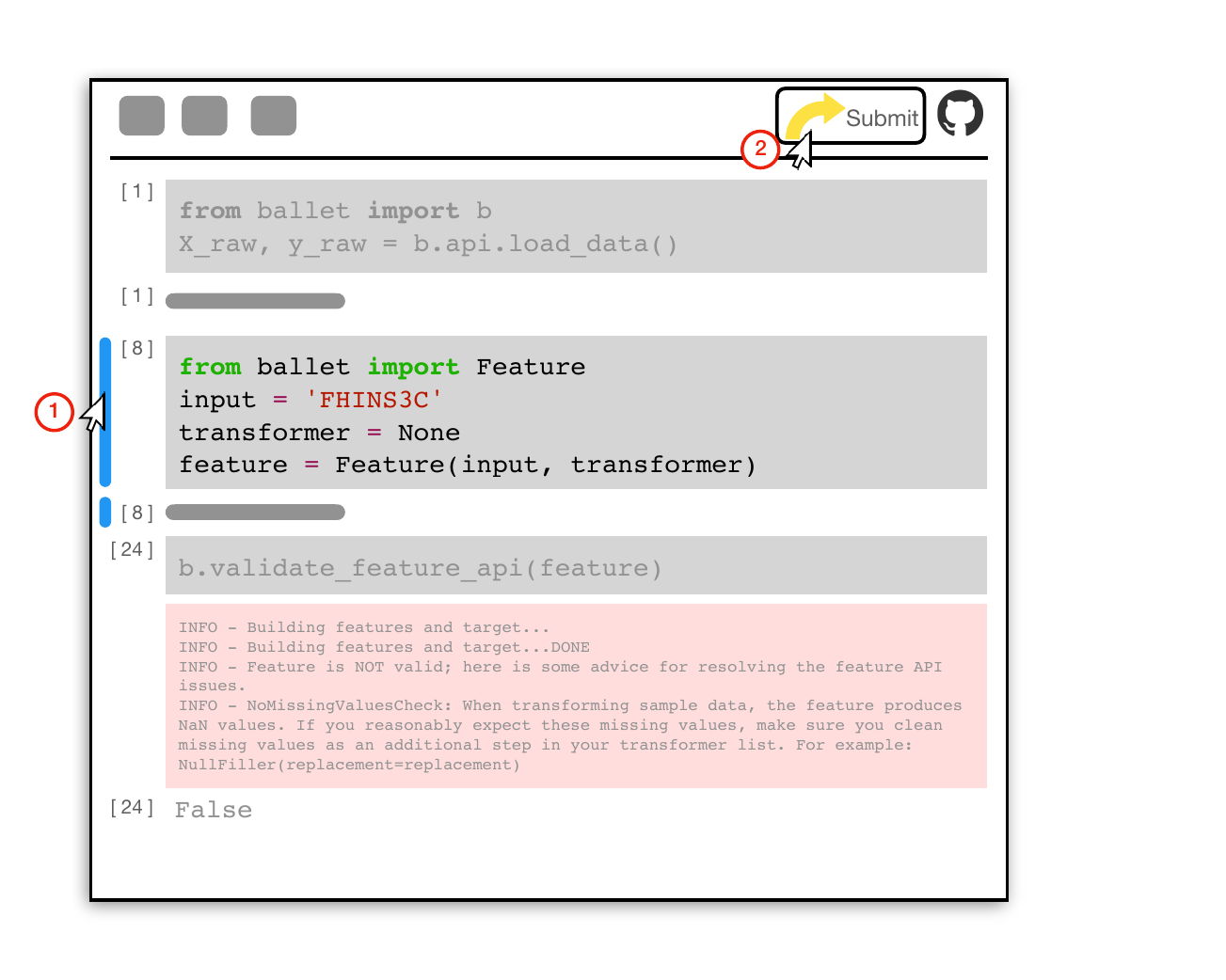}
    \quad
    \includegraphics[page=2, clip=true, trim={40 20 40 85}, width=0.45\textwidth]{ui}
    \caption{A overview of the \Assemble development environment. \Assemble{'s} frontend (left) extends JupyterLab to add a Submit button and a GitHub authentication button to the Notebook toolbar (top right). Users first authenticate \Assemble with GitHub using a supported OAuth flow. Then, after developing a patch within a larger, messy notebook, users select the code cell containing their desired patch using existing Notebook interactions (1) and press \Assemble{'s} Submit button (2) to cause it to be automatically formulated as a pull request by the backend. (2). The backend performs initial static analysis and validation of the intended submission and then creates a well-structured PR containing the patch (right). Taken together, the components of \Assemble support the \emph{patch contribution task} for notebook-based developers.}
    \Description{
        The left subfigure shows a stylized view of the \Assemble extension to the JupyterLab interface. There are three code cells, with execution counts 1, 8, and 24, respectively. The first code cell shows a code snippet for a developer to load sample data. The second code cell shows a code snippet of a developer using \Ballet{'s} feature development API to create a simple feature definition. This code cell is shown being selected by the developer using a mouse icon that is annotated to indicate ``step 1''. The third code cell shows a code snippet of a developer using \Ballet{'s} feature validation API to validate the feature definition from the previous code cell on sample data. The output is written to the standard error stream and shows that the feature failed its validation due to a failed ``NoMissingValuesCheck''. There are two icons in a toolbar on the top right. The first icon says ``Submit'' and is shown being selected by the developer using a mouse icon that is annotated to indicate ``step 2''. The second icon shows GitHub's logo.

        The right subfigure shows a stylized view of an open pull request containing the same code snippet that shown in the second code cell on the left subfigure, but this time the code snippet has been updated to fix the validation failure. The pull request introduces two new files at the file paths ``src/predict_x/features/contrib/user_bob/__init__.py'' and ``src/predict_x/features/contrib/user_bob/feature.py'', where the first file is empty but the second file contains the code snippet. The pull request is opened by a user ``bob'' and is asking to be pulled into the project ``ballet-predict-x'' of a different user ``alice''.
    }
    \label{fig:ui}
\end{figure}

In this paper, we describe the ideation, design, and implementation of a development environment that supports notebook-based collaborative data science. We then report on a user study of 23 developers using \Assemble in a data science collaboration case study. Finally, we discuss future directions for interfaces and workflows to support collaborative data science development.

\section{Background and Related Work}
\label{sec:related}

\subsection{Collaborative data science}

Researchers have observed that large collaborations in data science are rare compared to traditional software development, even as the field continues to grow and mature \cite{choi2017characteristics}. In data science development, one approach to collaboration is to coordinate around a shared work product, such as a data science pipeline \cite{crowston2019sociotechnical}. However the scale of previous collaborations has not approached what is regularly observed in open-source software development.

\Ballet is a framework that builds on approaches to facilitating collaboration between data scientists by carefully structuring a data science pipeline as a composition of smaller modules that can be developed separately \cite{smith2020enabling}. The creation of these small components can be done in parallel by different contributors who then introduce a patch to the upstream project defining this new component.

For example, as data science is an expansive and rapidly growing field, \Ballet focuses on predictive machine learning and an important subprocess called feature engineering, in which developers write code to transform raw variables into useful features suitable for learning algorithms. In such a feature engineering project, a feature engineering pipeline is composed of features; this pipeline is grown incrementally by patches/pull requests to the repository that contain \emph{feature definitions}. Thus developers might split up work to write individual feature definitions in short code snippets, in much the same way that they would split up work in fixing bugs or implementing new functionality in a traditional software project. The framework imposes a structure on the repository (i.e. feature definitions are written in individual Python source files at a specific path) and is then able to collect many feature definitions into a single feature engineering pipeline. It is also responsible for validating contributions using a custom suite of statistical and software tests. Other data scientists can freely install and use this collaboratively written feature engineering pipeline on their data that follows the same schema.

\Ballet leaves open the subsequent question of supporting data science developers in the process of creating feature definitions. One na\"{i}ve workflow that is available to software development experts is to use their usual low-level command line tools to manually structure their patches and make pull requests to the shared project repository \cite{gousios2014exploratory}. However, this workflow is challenging and discouraging for many data science developers. \Assemble fills this gap by allowing notebook-based developers to transparently contribute their work from within a notebook environment to a shared repository without exposing any low-level tools.

\subsection{Exploratory data analysis}

Similar to other data analysis tasks, creating a new feature definition is an iterative process in which data scientists explore patterns in the raw data, hypothesize about possible relationships, and evaluate potential transformations. Data scientists frequently use computational notebooks like Jupyter for exploratory analyses \cite{kery2017exploring, muller2019how}. However, functionality that is popular for exploratory work, such as interleaving code input and rich outputs and executing code cells in arbitrary orders, can can lead to analyses that are difficult to share, understand, reproduce, or productionize \cite{subramanian2020casual, chattopadhyay2020what, kery2018story}.

One response to this situation is to try to make the tools used in team-based software engineering more common in data science settings \cite{yang2018grounding, wilson2006software, amershi2019softwarea}, such as those for version control, testing, and building. These tools may be used more easily if they are embedded within the primary development environment, so both JupyterLab and other IDEs offer support for applying git operations, either natively or in extensions \cite{gitextension, gitplusextension}. However, the challenge of using git remains, albeit with a different visual interface. In contrast to these tools, while \Assemble is backed by git, it operates at a higher level of abstraction and its users never interact with git directly.

Researchers and practitioners have also built new tools that extend the functionality of existing tools to additional formats and settings. These include version control and provenance of notebooks and datasets, as well as other artifacts such as figures and learned model parameters; test frameworks for notebooks; and language extensions to support notebooks as libraries within a larger application \cite{pleban2020announcing, gori2020fileweaver}. 
Another approach is to attempt to improve the notebook experience itself with these issues in mind. For example, \citet{head2019managing} introduce code-gathering tools in a JupyterLab extension. With these tools, a notebook user can select certain content from a messy notebook, such as a variable assignment or displayed figure, and identify the set of code cells needed to produce that value along with the order in which they were executed. These can then be exported to a ``clean'' notebook. Other notebook extensions allow notebook users to track past analysis choices \cite{kery2019effective} or join with other developers for synchronous editing \cite{wang2019how}.

\subsection{Learning data science}

A separate line of related work inverts this problem and asks about challenges for non-ML experts learning ML. In the case of software developers, it is a lack of mathematical and theoretical background that causes hurdles \cite{cai2019software}, rather than difficulties using software development processes as in our case. For other non-experts, common pitfalls include challenges in formulating learning problems and over-reliance on headline performance metrics \cite{yang2018grounding}.

\section{Design}
\label{sec:design}

To investigate development workflow issues in \Ballet, we first conducted a formative study with eight data scientists recruited from a laboratory mailing list at a large research university. We asked them to write and submit feature definitions for a collaborative project based around predicting the incidence rates of dengue fever in two different regions. Although participants created feature definitions successfully, we observed that they struggled to contribute them to the shared repository using the pull request model, with only two creating a pull request at all. In interviews, participants acknowledged that a lack of familiarity and experience with the pull request-based model of open source development was an obstacle to contributing the code that they had written, especially in the context of team-based development \cite{gousios2015work}.

In this study, and in other experiments with \Ballet, we observed that data scientists predominately used notebooks to develop feature definitions before turning to entirely different environments and tools to extract the smallest relevant patch and create a pull request. We thus identified the \emph{patch contribution task} as an important interface problem to address in order to improve collaborative data science. Once working code has been written, we may be able to automate the entire process of code contribution according to the requirements of the specific project the user is working on.

With this in mind, we elicited the following design criteria to support patch contribution in a collaborative data science environment.

\begin{enumerate}
    \item[D1] \emph{Make code easy to contribute.} Once a patch has been identified, it should be easy to immediately contribute it without a separate process to productionize it.
    \item[D2] \emph{Hide low-level tools.} Unfamiliarity and difficulty with low-level tooling and processes, such as \inline{git} and the pull request model, tend to interrupt data scientists' ability to collaborate on a shared repository. Any solution to submitting patches should not include manual use of these tools.
    \item[D3] \emph{Minimize setup and installation friction.} Finally, the solution should fit seamlessly within users' existing development workflows, and should be easy to setup and install.
\end{enumerate}

Based on these criteria, we propose a design that extends the notebook interface to support submission of individual code cells as pull requests. By focusing on individual code cells, we allow developers to easily isolate relevant code to submit. Once a user has selected a code cell using existing Notebook interactions, pressing a simple, one-click ``Submit'' button added to the Notebook Toolbar panel spurs the creation and submission of a patch according to the configuration of the underlying project.

By abstracting away the low-level details of this process, we lose the ability to identify some code quality issues that would otherwise be identified by the tooling. To address this, we run an initial server-side validation using static analysis before forwarding on the patch, in order to immediately surface relevant problems to users within the notebook context. If submission is successful, the contributor can view their new PR in a matter of seconds. \Assemble is tightly integrated with Binder such that it can be launched from every \Ballet project via a README badge. Installation of the extension is handled automatically and the project settings are automatically detected so that developers can get right to work. An in-notebook, OAuth-based authentication flow also allows developers to easily authenticate with GitHub without difficult configuration.

In summary, we design \Assemble to provide the following functionality:

\begin{itemize}
    \item isolate relevant code snippets from a messy notebook
    \item transparently provide access to take actions on GitHub
    \item automatically formulate an isolated snippet as a PR to an upstream data science project without exposing any git details
\end{itemize}

\section{Implementation}
\label{sec:impl}

\begin{figure}
    \centering
    \includegraphics[page=1,width=0.48\textwidth]{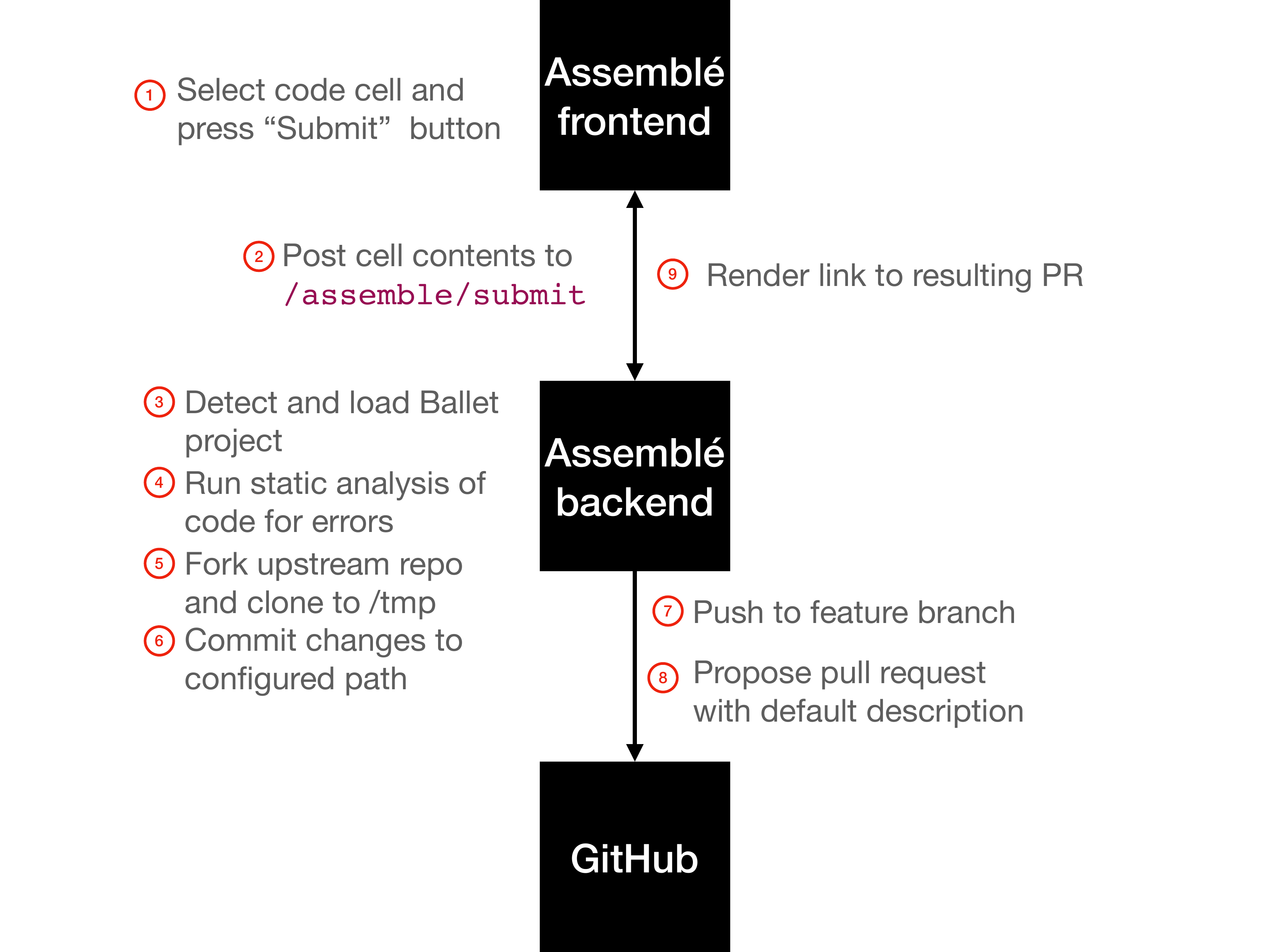}
    \includegraphics[page=2,width=0.48\textwidth]{implementation}
    \caption{The \Assemble environment includes both Jupyter frontend and server extensions. Completed code snippets are submitted directly to an upstream GitHub repository as pull requests (left). Users authenticate with GitHub using an accompanying OAuth proxy server (right).}
    \Description{
        The left subfigure shows an architecture diagram of \Assemble{'s} submit functionality. There are three nodes shown, ``\Assemble frontend'' (top), ``\Assemble backend'' (middle), and ``GitHub'' (bottom). Step 1 at \Assemble frontend: select code cell and press ``Submit'' button. Step 2 from \Assemble frontend to \Assemble backend: Post cell contents to ``/assemble/submit''. Steps 3, 4, 5, and 6 at \Assemble backend: Detect and load \Ballet project, run static analysis of code for errors, fork upstream repo and clone to ``/tmp'', commit changes to configured path. Steps 7 and 8 from \Assemble backend to GitHub: push to feature branch, propose pull request with default description. Step 9 from \Assemble backend to \Assemble frontend: render link to resulting PR.

        The right subfigure shows an architecture diagram of \Assemble{'s} GitHub authentication functionality. There are four nodes shown, ``\Assemble frontend'' (top left), ``\Assemble backend'' (bottom left), ``github-oauth-gateway'' (bottom right), and ``GitHub'' (top right). There are two workflows that occur in parallel, indicated by different colors. The first workflow starts with step 1 at \Assemble frontend: start OAuth flow. Step 2 at \Assemble backend: generate secret. Step 3 from \Assemble backend to GitHub: redirect to OAuth login page with secret state. Step 4 from GitHub to \Assemble frontend: request user credentials. Step 5 from GitHub to github-oauth-gateway: send access token and secret. Step 6 at github-oauth-gateway: store secret-token pair. The second workflow starts with step 1 from \Assemble frontend to \Assemble backend: prompt for token. Step 2 from \Assemble backend to github-oauth-gateway: poll for token associated with secret.
    }
    \label{fig:system}
    \end{figure}

\Assemble is implemented in three components: a JupyterLab frontend extension, a JupyterLab server extension, and an OAuth proxy server. These are shown in \Cref{fig:system}.

\subsection{JupyterLab extension}

The frontend extension is implemented in TypeScript on JupyterLab 2. It adds two buttons to the Notebook Panel toolbar. The GitHub button allows the user to initiate an authentication flow with GitHub (\Cref{sec:impl:github}). The Submit button identifies the currently selected code cell from the active notebook and extracts the source. It then posts the contents to the server to be submitted (D1). If the submission is successful, it displays a link to the GitHub pull request view. Otherwise, it shows a relevant error message -- usually a Python traceback due to syntax errors in the user's code.

The server extension is implemented in Python on Tornado 6. It adds routes to the Jupyter Server under the \inline{/assemble} prefix. These include \inline{/assemble/submit} to receive the code to be submitted, and three routes under \inline{/assemble/auth} to handle the authentication flow with GitHub. Upon extension initialization, it detects a Ballet project by ascending the file system, via the current working directory looking for the \inline{ballet.yml} file and loading the project using the \inline{ballet} library according to that configuration.

When the server extension receives the code to be submitted, it first runs a static analysis using Python's \inline{ast} module to ensure that it does not have syntax errors or undefined symbols, and automatically cleans/reformats the code to the target project's preferred style. It then prepares to submit it as a pull request. The upstream repository is determined from the project's settings and is forked, if needed, via the \inline{pygithub} interface to the GitHub API with the user's OAuth token, and cloned to a temporary directory. Using the Ballet client library, \Assemble can create an empty file at the correct path in the directory structure that will contain the proposed contribution, and writes to and commits this file. Depending on whether the user has contributed in the past, \Assemble may then also need to create additional files/folders to preserve the Python package structure (i.e. \inline{\_\_init\_\_.py} files). It then pushes to a new branch on the fork, and creates a pull request with a default description. Finally, it returns the pull request view link. This replaces what is usually 5-7 manual git operations with a robust and automated process (D2).

\subsection{GitHub authentication}
\label{sec:impl:github}

The final piece of the puzzle is authentication with GitHub, such that the server can act on GitHub as the user to create a new pull request. Most extensions that provide similar functionality (i.e. take some actions with an external service on behalf of a user that require authentication) ask the user to acquire a personal access token from the external service and provide it as a configuration variable, and in some cases register a web application using a developer console \cite{driveextension, githubextension}.

For our purposes, this is not acceptable, due to the high cost of setup for non-expert software developers (D2, D3). Instead, we would like to use OAuth \cite{oauth} to allow the user to enter their username and password for the service, and exchange them for a token that the server can use. However, this cannot be accomplished directly using the OAuth protocol because OAuth applications on GitHub (or elsewhere) must register a static callback URL. Instead, \Assemble might be running at any address, because with its Binder integration, the URLs assigned to Binder sessions are dynamic and on different domains.\footnote{For example, launching the same repository in a Binder can result in first a \inline{hub.gke.mybinder.org} URL and then an \inline{notebooks.gesis.org} URL, depending on the BinderHub deployment selected by the MyBinder load balancer.} To address this, we create \inline{github-oauth-gateway}, a lightweight proxy server for GitHub OAuth.\footnote{\url{\gatewayurl}} We create a reference deployment on Heroku and register it as an OAuth application with GitHub. Before the user can submit their code, they click the GitHub icon in the toolbar (\Cref{fig:ui}). This launches the OAuth flow. First the server creates a secret ``state'' at random. Then it redirects the user to the GitHub OAuth login page. The user is prompted to enter their username and password, and if the sign-in is successful, GitHub responds to the gateway with the token and the state created previously. The server polls the gateway for a token associated with its unique state, and receives the token in response when it is available.

\section{Evaluation}
\label{sec:evaluation}

We report our analysis of a user study in which 23 participants used \Assemble to develop and submit feature definitions to a shared repository as part of a collaborative project. We aim to assess the ability of users to successfully create pull requests for code snippets within a messy notebook and to identify key themes from participants' experiences.

\subsection{Procedures}

The \inline{predict-census-income} project\footnote{\url{\censusurl}} is a collaborative effort to predict personal income from responses to the U.S. Census American Community Survey (ACS). As part of a larger effort to understand collaborative data science practices \cite{smith2020enabling}, we recruited developers to use the \Ballet framework to develop and share feature definitions that would be predictive of personal income. In recruiting participants, we wanted to ensure that all participant backgrounds were represented: beginner/intermediate/expert developers in data science, software, and survey data analysis (the problem domain). Participants were entered into a drawing for several nominal prizes but were not otherwise compensated. In total, 27 developers were recruited after sampling personal contacts with a variety of backgrounds, posting to relevant mailing lists and message boards, and then using snowball sampling to reach more participants with similar backgrounds. Of these, 23 participants used \Assemble (v0.7.2) to develop their code and submit it to the shared repository.

Participants first completed a short questionnaire in which they self-reported their background in data science and open-source software development and their preferred development environments for data science tasks. Participants were also asked to consent to telemetry data collection. If they did, we instrumented \Assemble to collect detailed usage data on their development sessions, their use of the Submit button functionality, and their use of the \Ballet client library.

After completing their feature development, participants were also asked to take a short survey about their use of \Assemble, its features, and their overall experience with the project, and were invited to share free-response feedback.

We linked survey responses and telemetry data, and then removed all identifying information. Two researchers also qualitatively analyzed participants' free-response feedback using open and axial coding \cite{bohm2004theoretical}.

\subsection{Results}

Only five participants reported a preference for performing data science activities and Python development in notebook environments before the study, with 10 instead preferring IDEs and four preferring text editors. Seven participants had never contributed to open-source projects at all, while the remainder reported contributing approximately yearly (eight), monthly (four), or weekly (four). Fifteen participants opted into telemetry data collection, generating an average of 33 telemetry events each. A summary of participant background is shown in \Cref{fig:background}.

\subsection{Quantitative Result}
\begin{figure}
    \centering
    \subcaptionbox{Background of 23 developers using \Assemble in a study involving predicting personal income.\label{fig:results:background}}[0.55\textwidth]{\includegraphics[width=\linewidth]{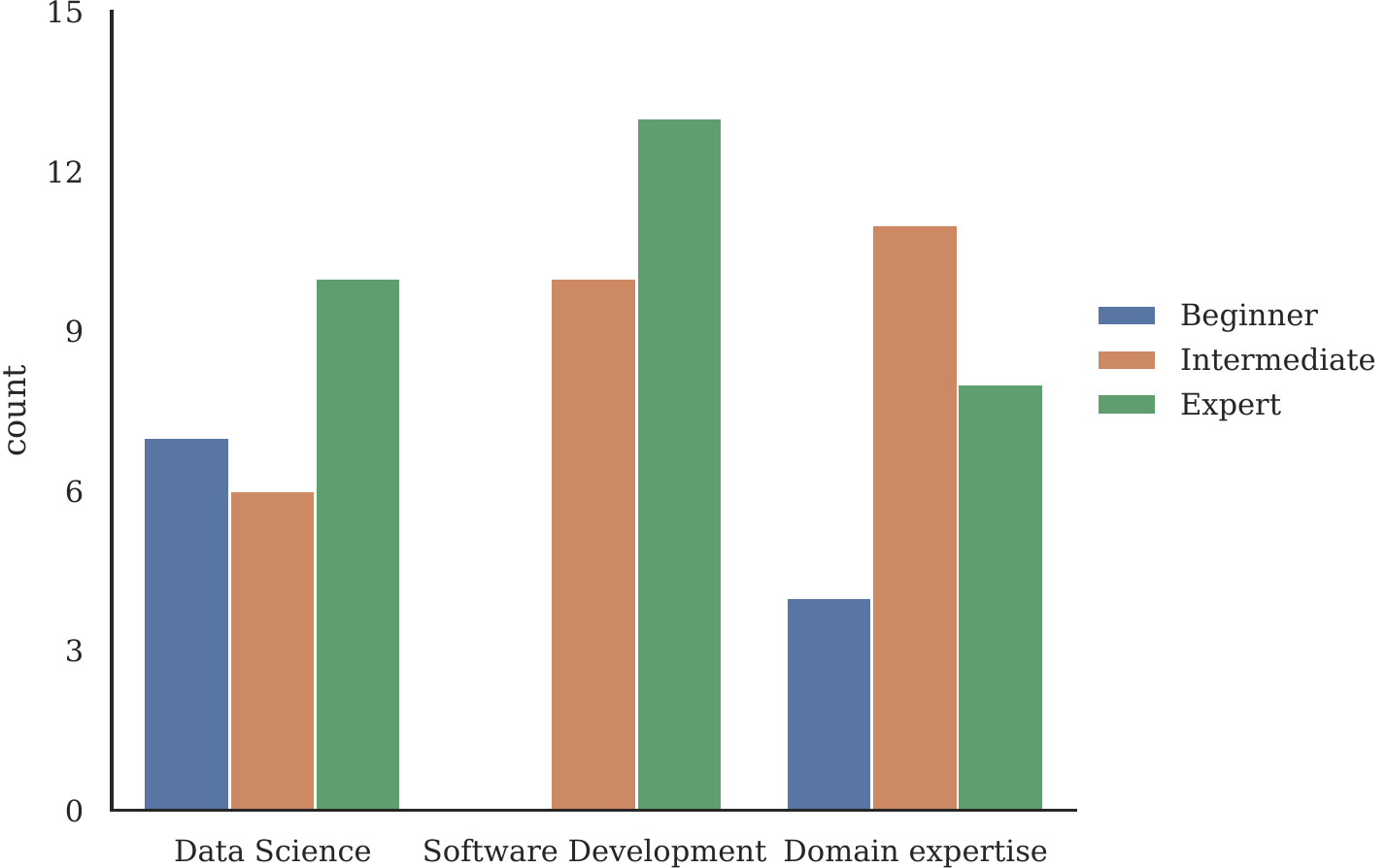}}
    \quad
    \subcaptionbox{Distribution of total minutes spent submitting features to upstream repository.\label{fig:results:minutes}}[0.38\textwidth]{\includegraphics[width=\linewidth]{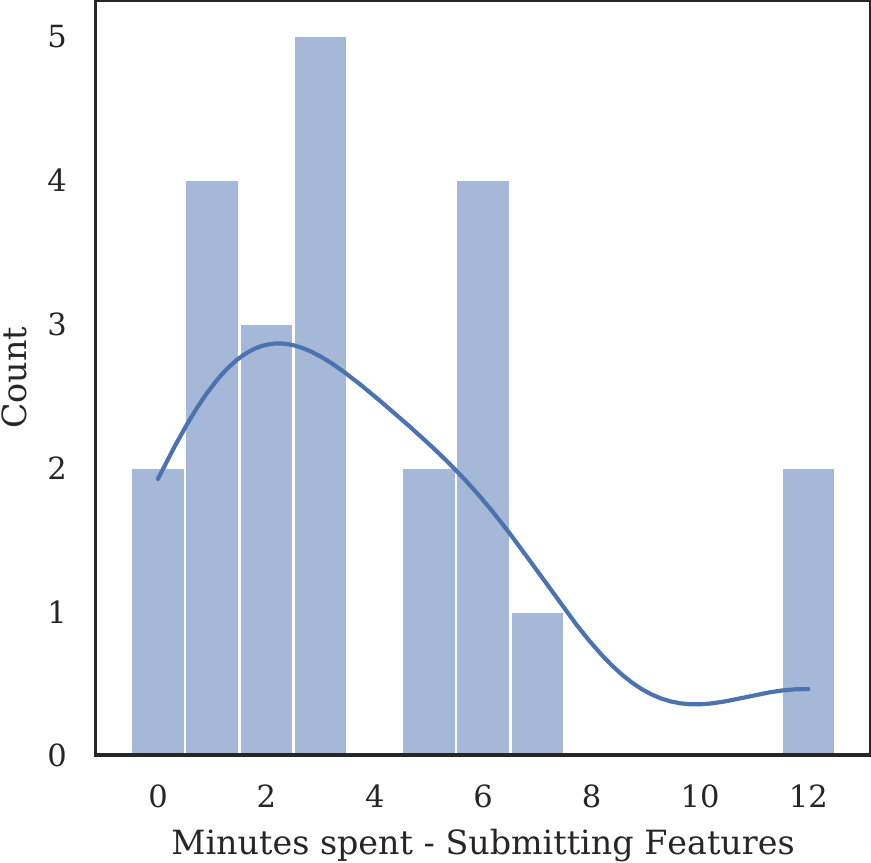}}
    \caption{User study results.}
    \Description{
        The left subfigure is a grouped bar chart with ``count'' on the y-axis. The first group shows data science beginners (7), intermediates (6), and experts (10). The second group shows software development beginners (0), intermediates (10), and experts (13). The third group shows domain expertise beginners (4), intermediates (11), and experts (8).

        The right subfigure is a histogram with a kernel density estimate (KDE). The x-axis shows ``minutes spent - submitting features'' and the x-axis shows ``count''. The bars of the histogram each cover a 1 minute interval and peak at 3 minutes, with the mode of the KDE at 2 minutes. About half of the mass of the distribution is between 0 and 3 minutes and the other half is between 3 minutes and 8 minutes. Then there are 2 outliers at 12 minutes.
    }
    \label{fig:background}
\end{figure}

Our main finding is that even with their diverse backgrounds and initial preferences, \emph{all participants in the study successfully used \Assemble to create one or more pull requests to the upstream project repository}. According to telemetry data, the modal user pressed the Submit button just once. Since the user study task was to submit a single feature, this suggests that users were immediately successful at creating a pull request for their desired contribution. We also find that participants were able to do this fairly quickly -- half were able to create a pull request using \Assemble in three minutes or less (\Cref{fig:results:minutes}). In 16 cases out of 45 submit events captured in the telemetry data (belonging to five unique users), \Assemble{'s} static analysis identified syntax errors in the intended submissions, each of which would have led to a pull request that would have failed \Ballet{'s} automated test suite. In all of these cases, users were able to quickly resolve these errors and submit again.

\subsection{Qualitative Results}

From a qualitative perspective, we identified two major themes from free text responses in our post-participation survey.

\paragraph{Keep It Simple While Introducing Better Affordances.}
Users overwhelmingly noted the simplicity with which they were able to submit their features, with one participant noting ``The process of integrating the new feature was very smooth" and another saying ``[Submitting a feature] was extremely and springily easy! Most rewarding part".
However, some participants noted that while the submission process was seamless, affordances could be better highlighted, e.g.: \emph{``Maybe highlight a bit more that you need to select the feature cell before hitting submit - I got confused after I missed this part "}.
Indeed, in the few cases where participants were not able to submit their feature on their first attempt, we see in our telemetry data that they either selected the wrong cell or introduced a syntax error.

\paragraph{Tensions between Abstraction and Submission Transparency. }
Submission transparency was another theme that emerged in our analysis. While we achieved our goal of hiding the lower-level procedures required in the pull request model, some participants were curious about the underlying process. Some wanted to know how their feature was evaluated, both in server-side validation and in continuous integration after pull request creation: ``It was not clear whether my feature was actually good, especially compared to other features." Others expressed a lack of understanding of what was going on "under the hood:" ``I didn't fully understand how \Assemble was working on the backend to actually develop the feature with relatively straightforward commands, but it seemed to work pretty well."
This feedback highlights the tension between abstraction and transparency. While users clearly appreciated the simplicity facilitated by the submission mechanism, they missed the traceability and feedback a more classical pull request model would have provided. We see this as an opportunity to introduce optionally available traces detailing the steps of the underlying process, partly as a way of onboarding non-experts into the open-source development workflow.

\section{Discussion and Conclusion}
\label{sec:conclusion}

For developers to select the code to submit, we rely upon simple existing interactions provided by Jupyter (i.e. select one cell, select multiple cells). However, as some developers requested \textit{better affordances}, other interactions could be incorporated. For example, code gathering tools \cite{head2019managing} could provide the means to more easily select the code to be submitted, while staying within the notebook environment.

One reason (of many) that powerful developer tools exist for team-based version control is to avoid and/or resolve merge conflicts when contributions are often scattered across multiple files. By tightly coupling with \Ballet projects, we take advantage of the structure those projects impose on contributions: they are in single Python modules at well-defined locations in the directory structure. By relaxing this assumption, or by defining such structures for other settings, \Assemble could be used more generally to contribute patches to central locations. For example, the ML Bazaar framework \cite{smith2020machine} organizes data science pipelines into a graph of ``ML primitives,'' each of which requires experts to create and validate the primitive (a JSON annotation of some underlying implementation) in a notebook. \Assemble could be used to extract the completed primitive and submit it to the project's curated catalog of community primitives. As another example, an educator running an introductory programming class could invite students to submit their implementations of a basic algorithm to a joint hosting repository, such that they could share in the code review process and learn from the implementations of others. Similarly, a Python language extension for sharing simple functions \cite{fast2016meta} could use the functionality of the development environment to share functions, rather than requiring the manual addition of function decorators.

In this paper, we presented the design and implementation of the \Assemble development environment. We showed how this environment vastly improves the efficiency and experience of developers collaborating on a data science pipeline using the \Ballet framework, and how the use of JupyterLab extensions allows developers to share code without leaving the notebook.

\section*{Acknowledgment}

We'd like to thank the participants of the ballet-predict-census-income study. This work is supported in part by NSF Award 1761812.

\bibliographystyle{ACM-Reference-Format}
\balance
\bibliography{references}


\begin{thebibliography}{27}


\ifx \showCODEN    \undefined \def \showCODEN     #1{\unskip}     \fi
\ifx \showDOI      \undefined \def \showDOI       #1{#1}\fi
\ifx \showISBNx    \undefined \def \showISBNx     #1{\unskip}     \fi
\ifx \showISBNxiii \undefined \def \showISBNxiii  #1{\unskip}     \fi
\ifx \showISSN     \undefined \def \showISSN      #1{\unskip}     \fi
\ifx \showLCCN     \undefined \def \showLCCN      #1{\unskip}     \fi
\ifx \shownote     \undefined \def \shownote      #1{#1}          \fi
\ifx \showarticletitle \undefined \def \showarticletitle #1{#1}   \fi
\ifx \showURL      \undefined \def \showURL       {\relax}        \fi
\providecommand\bibfield[2]{#2}
\providecommand\bibinfo[2]{#2}
\providecommand\natexlab[1]{#1}
\providecommand\showeprint[2][]{arXiv:#2}

\bibitem[\protect\citeauthoryear{Amershi, Begel, Bird, DeLine, Gall, Kamar,
  Nagappan, Nushi, and Zimmermann}{Amershi et~al\mbox{.}}{2019}]%
        {amershi2019softwarea}
\bibfield{author}{\bibinfo{person}{Saleema Amershi}, \bibinfo{person}{Andrew
  Begel}, \bibinfo{person}{Christian Bird}, \bibinfo{person}{Robert DeLine},
  \bibinfo{person}{Harald Gall}, \bibinfo{person}{Ece Kamar},
  \bibinfo{person}{Nachiappan Nagappan}, \bibinfo{person}{Besmira Nushi}, {and}
  \bibinfo{person}{Thomas Zimmermann}.} \bibinfo{year}{2019}\natexlab{}.
\newblock \showarticletitle{Software {{Engineering}} for {{Machine Learning}}:
  {{A Case Study}}}. In \bibinfo{booktitle}{\emph{2019 {{IEEE}}/{{ACM}} 41st
  {{International Conference}} on {{Software Engineering}}: {{Software
  Engineering}} in {{Practice}} ({{ICSE}}-{{SEIP}})}}.
  \bibinfo{publisher}{{IEEE}}, \bibinfo{address}{{Montreal, QC, Canada}},
  \bibinfo{pages}{291--300}.
\newblock
\showISBNx{978-1-72811-760-7}
\urldef\tempurl%
\url{https://doi.org/10.1109/ICSE-SEIP.2019.00042}
\showDOI{\tempurl}


\bibitem[\protect\citeauthoryear{B{\"o}hm}{B{\"o}hm}{2004}]%
        {bohm2004theoretical}
\bibfield{author}{\bibinfo{person}{Andreas B{\"o}hm}.}
  \bibinfo{year}{2004}\natexlab{}.
\newblock \showarticletitle{Theoretical Coding: Text Analysis in}.
\newblock \bibinfo{journal}{\emph{A companion to qualitative research}}
  \bibinfo{volume}{1} (\bibinfo{year}{2004}).
\newblock


\bibitem[\protect\citeauthoryear{Cai and Guo}{Cai and Guo}{2019}]%
        {cai2019software}
\bibfield{author}{\bibinfo{person}{Carrie~J. Cai} {and}
  \bibinfo{person}{Philip~J. Guo}.} \bibinfo{year}{2019}\natexlab{}.
\newblock \showarticletitle{Software {{Developers Learning Machine Learning}}:
  {{Motivations}}, {{Hurdles}}, and {{Desires}}}. In
  \bibinfo{booktitle}{\emph{2019 {{IEEE Symposium}} on {{Visual Languages}} and
  {{Human}}-{{Centric Computing}} ({{VL}}/{{HCC}})}}.
  \bibinfo{publisher}{{IEEE}}, \bibinfo{address}{{Memphis, TN, USA}},
  \bibinfo{pages}{25--34}.
\newblock
\showISBNx{978-1-72810-810-0}
\urldef\tempurl%
\url{https://doi.org/10.1109/VLHCC.2019.8818751}
\showDOI{\tempurl}


\bibitem[\protect\citeauthoryear{Chattopadhyay, Prasad, Henley, Sarma, and
  Barik}{Chattopadhyay et~al\mbox{.}}{2020}]%
        {chattopadhyay2020what}
\bibfield{author}{\bibinfo{person}{Souti Chattopadhyay},
  \bibinfo{person}{Ishita Prasad}, \bibinfo{person}{Austin~Z. Henley},
  \bibinfo{person}{Anita Sarma}, {and} \bibinfo{person}{Titus Barik}.}
  \bibinfo{year}{2020}\natexlab{}.
\newblock \showarticletitle{What's {{Wrong}} with {{Computational Notebooks}}?
  {{Pain Points}}, {{Needs}}, and {{Design Opportunities}}}. In
  \bibinfo{booktitle}{\emph{Proceedings of the 2020 {{CHI Conference}} on
  {{Human Factors}} in {{Computing Systems}}}}. \bibinfo{publisher}{{ACM}},
  \bibinfo{address}{{Honolulu HI USA}}, \bibinfo{pages}{1--12}.
\newblock
\showISBNx{978-1-4503-6708-0}
\urldef\tempurl%
\url{https://doi.org/10.1145/3313831.3376729}
\showDOI{\tempurl}


\bibitem[\protect\citeauthoryear{Choi and Tausczik}{Choi and Tausczik}{2017}]%
        {choi2017characteristics}
\bibfield{author}{\bibinfo{person}{Joohee Choi} {and} \bibinfo{person}{Yla
  Tausczik}.} \bibinfo{year}{2017}\natexlab{}.
\newblock \showarticletitle{Characteristics of Collaboration in the Emerging
  Practice of Open Data Analysis}. \bibinfo{publisher}{ACM Press},
  \bibinfo{pages}{835–846}.
\newblock
\showISBNx{978-1-4503-4335-0}
\urldef\tempurl%
\url{https://doi.org/10.1145/2998181.2998265}
\showDOI{\tempurl}


\bibitem[\protect\citeauthoryear{Crowston, Saltz, Rezgui, Hegde, and
  You}{Crowston et~al\mbox{.}}{2019}]%
        {crowston2019sociotechnical}
\bibfield{author}{\bibinfo{person}{Kevin Crowston}, \bibinfo{person}{Jeff~S.
  Saltz}, \bibinfo{person}{Amira Rezgui}, \bibinfo{person}{Yatish Hegde}, {and}
  \bibinfo{person}{Sangseok You}.} \bibinfo{year}{2019}\natexlab{}.
\newblock \showarticletitle{Socio-Technical {{Affordances}} for {{Stigmergic
  Coordination Implemented}} in {{MIDST}}, a {{Tool}} for {{Data}}-{{Science
  Teams}}}.
\newblock \bibinfo{journal}{\emph{Proceedings of the ACM on Human-Computer
  Interaction}} \bibinfo{volume}{3}, \bibinfo{number}{CSCW}
  (\bibinfo{date}{Nov.} \bibinfo{year}{2019}), \bibinfo{pages}{1--25}.
\newblock
\showISSN{2573-0142, 2573-0142}
\urldef\tempurl%
\url{https://doi.org/10.1145/3359219}
\showDOI{\tempurl}


\bibitem[\protect\citeauthoryear{Fast and Bernstein}{Fast and
  Bernstein}{2016}]%
        {fast2016meta}
\bibfield{author}{\bibinfo{person}{Ethan Fast} {and}
  \bibinfo{person}{Michael~S. Bernstein}.} \bibinfo{year}{2016}\natexlab{}.
\newblock \showarticletitle{Meta: Enabling Programming Languages to Learn from
  the Crowd}. In \bibinfo{booktitle}{\emph{Proceedings of the 29th Annual
  Symposium on User Interface Software and Technology - UIST ’16}}.
  \bibinfo{publisher}{ACM Press}, \bibinfo{pages}{259–270}.
\newblock
\showISBNx{978-1-4503-4189-9}
\urldef\tempurl%
\url{https://doi.org/10.1145/2984511.2984532}
\showDOI{\tempurl}


\bibitem[\protect\citeauthoryear{Gori, Han, and {Beaudouin-Lafon}}{Gori
  et~al\mbox{.}}{2020}]%
        {gori2020fileweaver}
\bibfield{author}{\bibinfo{person}{Julien Gori}, \bibinfo{person}{Han~L. Han},
  {and} \bibinfo{person}{Michel {Beaudouin-Lafon}}.}
  \bibinfo{year}{2020}\natexlab{}.
\newblock \showarticletitle{{{FileWeaver}}: {{Flexible File Management}} with
  {{Automatic Dependency Tracking}}}. In \bibinfo{booktitle}{\emph{Proceedings
  of the 33rd {{Annual ACM Symposium}} on {{User Interface Software}} and
  {{Technology}}}}. \bibinfo{publisher}{{ACM}}, \bibinfo{address}{{Virtual
  Event USA}}, \bibinfo{pages}{22--34}.
\newblock
\showISBNx{978-1-4503-7514-6}
\urldef\tempurl%
\url{https://doi.org/10.1145/3379337.3415830}
\showDOI{\tempurl}


\bibitem[\protect\citeauthoryear{Gousios, Pinzger, and van Deursen}{Gousios
  et~al\mbox{.}}{2014}]%
        {gousios2014exploratory}
\bibfield{author}{\bibinfo{person}{Georgios Gousios}, \bibinfo{person}{Martin
  Pinzger}, {and} \bibinfo{person}{Arie van Deursen}.}
  \bibinfo{year}{2014}\natexlab{}.
\newblock \showarticletitle{An Exploratory Study of the Pull-Based Software
  Development Model}. In \bibinfo{booktitle}{\emph{Proceedings of the 36th
  {{International Conference}} on {{Software Engineering}} - {{ICSE}} 2014}}.
  \bibinfo{publisher}{{ACM Press}}, \bibinfo{address}{{Hyderabad, India}},
  \bibinfo{pages}{345--355}.
\newblock
\showISBNx{978-1-4503-2756-5}
\urldef\tempurl%
\url{https://doi.org/10.1145/2568225.2568260}
\showDOI{\tempurl}


\bibitem[\protect\citeauthoryear{Gousios, Zaidman, Storey, and
  Van~Deursen}{Gousios et~al\mbox{.}}{2015}]%
        {gousios2015work}
\bibfield{author}{\bibinfo{person}{Georgios Gousios}, \bibinfo{person}{Andy
  Zaidman}, \bibinfo{person}{Margaret-Anne Storey}, {and} \bibinfo{person}{Arie
  Van~Deursen}.} \bibinfo{year}{2015}\natexlab{}.
\newblock \showarticletitle{Work practices and challenges in pull-based
  development: the integrator's perspective}. In \bibinfo{booktitle}{\emph{2015
  IEEE/ACM 37th IEEE International Conference on Software Engineering}},
  Vol.~\bibinfo{volume}{1}. IEEE, \bibinfo{pages}{358--368}.
\newblock


\bibitem[\protect\citeauthoryear{Head, Hohman, Barik, Drucker, and DeLine}{Head
  et~al\mbox{.}}{2019}]%
        {head2019managing}
\bibfield{author}{\bibinfo{person}{Andrew Head}, \bibinfo{person}{Fred Hohman},
  \bibinfo{person}{Titus Barik}, \bibinfo{person}{Steven~M. Drucker}, {and}
  \bibinfo{person}{Robert DeLine}.} \bibinfo{year}{2019}\natexlab{}.
\newblock \showarticletitle{Managing Messes in Computational Notebooks}. In
  \bibinfo{booktitle}{\emph{Proceedings of the 2019 CHI Conference on Human
  Factors in Computing Systems}} (Glasgow, Scotland Uk)
  \emph{(\bibinfo{series}{CHI '19})}. \bibinfo{publisher}{ACM},
  \bibinfo{address}{New York, NY, USA}, Article \bibinfo{articleno}{270},
  \bibinfo{numpages}{12}~pages.
\newblock


\bibitem[\protect\citeauthoryear{Kery, John, O'Flaherty, Horvath, and
  Myers}{Kery et~al\mbox{.}}{2019}]%
        {kery2019effective}
\bibfield{author}{\bibinfo{person}{Mary~Beth Kery}, \bibinfo{person}{Bonnie~E.
  John}, \bibinfo{person}{Patrick O'Flaherty}, \bibinfo{person}{Amber Horvath},
  {and} \bibinfo{person}{Brad~A. Myers}.} \bibinfo{year}{2019}\natexlab{}.
\newblock \showarticletitle{Towards {{Effective Foraging}} by {{Data
  Scientists}} to {{Find Past Analysis Choices}}}. In
  \bibinfo{booktitle}{\emph{Proceedings of the 2019 {{CHI Conference}} on
  {{Human Factors}} in {{Computing Systems}}}}. \bibinfo{publisher}{{ACM}},
  \bibinfo{address}{{Glasgow Scotland Uk}}, \bibinfo{pages}{1--13}.
\newblock
\showISBNx{978-1-4503-5970-2}
\urldef\tempurl%
\url{https://doi.org/10.1145/3290605.3300322}
\showDOI{\tempurl}


\bibitem[\protect\citeauthoryear{Kery and Myers}{Kery and Myers}{2017}]%
        {kery2017exploring}
\bibfield{author}{\bibinfo{person}{Mary~Beth Kery} {and}
  \bibinfo{person}{Brad~A. Myers}.} \bibinfo{year}{2017}\natexlab{}.
\newblock \showarticletitle{Exploring exploratory programming}. In
  \bibinfo{booktitle}{\emph{2017 IEEE Symposium on Visual Languages and
  Human-Centric Computing (VL/HCC)}}. \bibinfo{pages}{25–29}.
\newblock
\showISSN{1943-6106}
\urldef\tempurl%
\url{https://doi.org/10.1109/VLHCC.2017.8103446}
\showDOI{\tempurl}


\bibitem[\protect\citeauthoryear{Kery, Radensky, Arya, John, and Myers}{Kery
  et~al\mbox{.}}{2018}]%
        {kery2018story}
\bibfield{author}{\bibinfo{person}{Mary~Beth Kery}, \bibinfo{person}{Marissa
  Radensky}, \bibinfo{person}{Mahima Arya}, \bibinfo{person}{Bonnie~E. John},
  {and} \bibinfo{person}{Brad~A. Myers}.} \bibinfo{year}{2018}\natexlab{}.
\newblock \showarticletitle{The {{Story}} in the {{Notebook}}: {{Exploratory
  Data Science}} Using a {{Literate Programming Tool}}}. In
  \bibinfo{booktitle}{\emph{Proceedings of the 2018 {{CHI Conference}} on
  {{Human Factors}} in {{Computing Systems}}}} \emph{(\bibinfo{series}{{{CHI}}
  '18})}. \bibinfo{publisher}{{Association for Computing Machinery}},
  \bibinfo{address}{{New York, NY, USA}}, \bibinfo{pages}{1--11}.
\newblock
\showISBNx{978-1-4503-5620-6}
\urldef\tempurl%
\url{https://doi.org/10.1145/3173574.3173748}
\showDOI{\tempurl}


\bibitem[\protect\citeauthoryear{Muller, Lange, Wang, Piorkowski, Tsay, Liao,
  Dugan, and Erickson}{Muller et~al\mbox{.}}{2019}]%
        {muller2019how}
\bibfield{author}{\bibinfo{person}{Michael Muller}, \bibinfo{person}{Ingrid
  Lange}, \bibinfo{person}{Dakuo Wang}, \bibinfo{person}{David Piorkowski},
  \bibinfo{person}{Jason Tsay}, \bibinfo{person}{Q.~Vera Liao},
  \bibinfo{person}{Casey Dugan}, {and} \bibinfo{person}{Thomas Erickson}.}
  \bibinfo{year}{2019}\natexlab{}.
\newblock \showarticletitle{How {{Data Science Workers Work}} with {{Data}}:
  {{Discovery}}, {{Capture}}, {{Curation}}, {{Design}}, {{Creation}}}. In
  \bibinfo{booktitle}{\emph{Proceedings of the 2019 {{CHI Conference}} on
  {{Human Factors}} in {{Computing Systems}}}} \emph{(\bibinfo{series}{{{CHI}}
  '19})}. \bibinfo{publisher}{{Association for Computing Machinery}},
  \bibinfo{address}{{New York, NY, USA}}, \bibinfo{pages}{1--15}.
\newblock
\showISBNx{978-1-4503-5970-2}
\urldef\tempurl%
\url{https://doi.org/10.1145/3290605.3300356}
\showDOI{\tempurl}


\bibitem[\protect\citeauthoryear{{OAuth Working Group}}{{OAuth Working
  Group}}{2012}]%
        {oauth}
\bibfield{author}{\bibinfo{person}{{OAuth Working Group}}.}
  \bibinfo{year}{2012}\natexlab{}.
\newblock \bibinfo{booktitle}{\emph{The OAuth 2.0 Authorization Framework}}.
\newblock \bibinfo{type}{RFC} 6749. \bibinfo{institution}{RFC Editor}.
\newblock
\urldef\tempurl%
\url{https://www.rfc-editor.org/rfc/rfc6749}
\showURL{%
\tempurl}


\bibitem[\protect\citeauthoryear{Pleban}{Pleban}{2020}]%
        {pleban2020announcing}
\bibfield{author}{\bibinfo{person}{Dean Pleban}.}
  \bibinfo{year}{2020}\natexlab{}.
\newblock \bibinfo{title}{Announcing {{Data Science Pull Requests}}}.
\newblock
  \bibinfo{howpublished}{https://dagshub.com/blog/data-science-pull-requests/}.
\newblock


\bibitem[\protect\citeauthoryear{{Project Jupyter Contributors}}{{Project
  Jupyter Contributors}}{[n.d.]a}]%
        {gitextension}
\bibfield{author}{\bibinfo{person}{{Project Jupyter Contributors}}.}
  \bibinfo{year}{[n.d.]}\natexlab{a}.
\newblock \bibinfo{title}{JupyterLab Git}.
\newblock
  \bibinfo{howpublished}{\url{https://github.com/jupyterlab/jupyterlab-git}}.
\newblock
\newblock
\shownote{Accessed on 2021-03-24 (commit 267c149).}


\bibitem[\protect\citeauthoryear{{Project Jupyter Contributors}}{{Project
  Jupyter Contributors}}{[n.d.]b}]%
        {githubextension}
\bibfield{author}{\bibinfo{person}{{Project Jupyter Contributors}}.}
  \bibinfo{year}{[n.d.]}\natexlab{b}.
\newblock \bibinfo{title}{JupyterLab GitHub}.
\newblock
  \bibinfo{howpublished}{\url{https://github.com/jupyterlab/jupyterlab-github}}.
\newblock
\newblock
\shownote{Accessed on 2021-01-10 (commit 065aa44).}


\bibitem[\protect\citeauthoryear{{Project Jupyter Contributors}}{{Project
  Jupyter Contributors}}{[n.d.]c}]%
        {driveextension}
\bibfield{author}{\bibinfo{person}{{Project Jupyter Contributors}}.}
  \bibinfo{year}{[n.d.]}\natexlab{c}.
\newblock \bibinfo{title}{jupyterlab-google-drive}.
\newblock
  \bibinfo{howpublished}{\url{https://github.com/jupyterlab/jupyterlab-google-drive}}.
\newblock
\newblock
\shownote{Accessed on 2021-01-10 (commit ab727c4).}


\bibitem[\protect\citeauthoryear{Rathi}{Rathi}{[n.d.]}]%
        {gitplusextension}
\bibfield{author}{\bibinfo{person}{Amit Rathi}.}
  \bibinfo{year}{[n.d.]}\natexlab{}.
\newblock \bibinfo{title}{JupyterLab GitPlus}.
\newblock
  \bibinfo{howpublished}{\url{https://github.com/ReviewNB/jupyterlab-gitplus}}.
\newblock
\newblock
\shownote{Accessed on 2021-03-24 (commit c6cfa76).}


\bibitem[\protect\citeauthoryear{Smith, Cito, Lu, and Veeramachaneni}{Smith
  et~al\mbox{.}}{2020a}]%
        {smith2020enabling}
\bibfield{author}{\bibinfo{person}{Micah~J. Smith}, \bibinfo{person}{J{\"u}rgen
  Cito}, \bibinfo{person}{Kelvin Lu}, {and} \bibinfo{person}{Kalyan
  Veeramachaneni}.} \bibinfo{year}{2020}\natexlab{a}.
\newblock \showarticletitle{Enabling Collaborative Data Science Development
  with the {{Ballet}} Framework}.
\newblock \bibinfo{journal}{\emph{arXiv:2012.07816 [cs]}}
  (\bibinfo{date}{December} \bibinfo{year}{2020}).
\newblock
\showeprint[arxiv]{2012.07816}~[cs]


\bibitem[\protect\citeauthoryear{Smith, Sala, Kanter, and Veeramachaneni}{Smith
  et~al\mbox{.}}{2020b}]%
        {smith2020machine}
\bibfield{author}{\bibinfo{person}{Micah~J. Smith}, \bibinfo{person}{Carles
  Sala}, \bibinfo{person}{James~Max Kanter}, {and} \bibinfo{person}{Kalyan
  Veeramachaneni}.} \bibinfo{year}{2020}\natexlab{b}.
\newblock \showarticletitle{The {{Machine Learning Bazaar}}: {{Harnessing}} the
  {{ML Ecosystem}} for {{Effective System Development}}}. In
  \bibinfo{booktitle}{\emph{Proceedings of the 2020 {{ACM SIGMOD International
  Conference}} on {{Management}} of {{Data}}}}
  \emph{(\bibinfo{series}{{{SIGMOD}} '20})}. \bibinfo{publisher}{{Association
  for Computing Machinery}}, \bibinfo{address}{{Portland, OR, USA}},
  \bibinfo{pages}{785--800}.
\newblock
\showISBNx{978-1-4503-6735-6}
\urldef\tempurl%
\url{https://doi.org/10.1145/3318464.3386146}
\showDOI{\tempurl}


\bibitem[\protect\citeauthoryear{Subramanian, Hamdan, and Borchers}{Subramanian
  et~al\mbox{.}}{2020}]%
        {subramanian2020casual}
\bibfield{author}{\bibinfo{person}{Krishna Subramanian}, \bibinfo{person}{Nur
  Hamdan}, {and} \bibinfo{person}{Jan Borchers}.}
  \bibinfo{year}{2020}\natexlab{}.
\newblock \showarticletitle{Casual Notebooks and Rigid Scripts: Understanding
  Data Science Programming}. In \bibinfo{booktitle}{\emph{2020 IEEE Symposium
  on Visual Languages and Human-Centric Computing (VL/HCC)}}.
  \bibinfo{pages}{1–5}.
\newblock
\showISSN{1943-6106}
\urldef\tempurl%
\url{https://doi.org/10.1109/VL/HCC50065.2020.9127207}
\showDOI{\tempurl}


\bibitem[\protect\citeauthoryear{Wang, Mittal, Brooks, and Oney}{Wang
  et~al\mbox{.}}{2019}]%
        {wang2019how}
\bibfield{author}{\bibinfo{person}{April~Yi Wang}, \bibinfo{person}{Anant
  Mittal}, \bibinfo{person}{Christopher Brooks}, {and} \bibinfo{person}{Steve
  Oney}.} \bibinfo{year}{2019}\natexlab{}.
\newblock \showarticletitle{How {{Data Scientists Use Computational Notebooks}}
  for {{Real}}-{{Time Collaboration}}}.
\newblock \bibinfo{journal}{\emph{Proceedings of the ACM on Human-Computer
  Interaction}} \bibinfo{volume}{3}, \bibinfo{number}{CSCW}
  (\bibinfo{date}{Nov.} \bibinfo{year}{2019}), \bibinfo{pages}{1--30}.
\newblock
\showISSN{2573-0142}
\urldef\tempurl%
\url{https://doi.org/10.1145/3359141}
\showDOI{\tempurl}


\bibitem[\protect\citeauthoryear{Wilson}{Wilson}{2006}]%
        {wilson2006software}
\bibfield{author}{\bibinfo{person}{G. Wilson}.}
  \bibinfo{year}{2006}\natexlab{}.
\newblock \showarticletitle{Software {{Carpentry}}: {{Getting Scientists}} to
  {{Write Better Code}} by {{Making Them More Productive}}}.
\newblock \bibinfo{journal}{\emph{Computing in Science \& Engineering}}
  \bibinfo{volume}{8}, \bibinfo{number}{6} (\bibinfo{date}{Nov.}
  \bibinfo{year}{2006}), \bibinfo{pages}{66--69}.
\newblock
\showISSN{1521-9615}
\urldef\tempurl%
\url{https://doi.org/10.1109/MCSE.2006.122}
\showDOI{\tempurl}


\bibitem[\protect\citeauthoryear{Yang, Suh, Chen, and Ramos}{Yang
  et~al\mbox{.}}{2018}]%
        {yang2018grounding}
\bibfield{author}{\bibinfo{person}{Qian Yang}, \bibinfo{person}{Jina Suh},
  \bibinfo{person}{Nan-Chen Chen}, {and} \bibinfo{person}{Gonzalo Ramos}.}
  \bibinfo{year}{2018}\natexlab{}.
\newblock \showarticletitle{Grounding {{Interactive Machine Learning Tool
  Design}} in {{How Non}}-{{Experts Actually Build Models}}}.
\newblock \bibinfo{journal}{\emph{Proceedings of the 2018 on Designing
  Interactive Systems Conference 2018 - DIS '18}} (\bibinfo{year}{2018}),
  \bibinfo{pages}{573--584}.
\newblock
\showISBNx{9781450351980}
\urldef\tempurl%
\url{https://doi.org/10.1145/3196709.3196729}
\showDOI{\tempurl}


\end{thebibliography}

\end{document}